\documentclass[pra,reprint, twocolumn, groupedaddress, showpacs]{revtex4}
\usepackage{amsmath}
\usepackage{graphicx}

\begin{document}


\title{Coherent Excitation of the 6S$_{1/2}$ to 5D$_{3/2}$ Electric Quadrupole Transition in $^{138}$Ba$^+$}



\author{Adam Kleczewski}
\author{Matthew Hoffman}
\email[]{hoffman2@u.washington.edu}
\author{J.A. Sherman}
\altaffiliation{Present address: NIST-Boulder / Optical Frequency Measurements group,
Boulder, CO 80305-3337}
\author{Eric Magnuson}
\altaffiliation{Present address: Department of Physics,
University of Virginia,
Charlottesville, VA 22901}
\author{Boris B. Blinov}
\author{E. N. Fortson}
\affiliation{University of Washington, Department of Physics, Seattle, Washington 98195, USA}


\date{\today}

\begin{abstract}
The electric dipole-forbidden, quadrupole 6S$_{1/2} \leftrightarrow$ 5D$_{3/2}$ transition in Ba$^{+}$ near 2051 nm, with a natural linewidth of 13 mHz, is attractive for potential observation of parity non-conservation, and also as a clock transition for a barium ion optical frequency standard.  This transition also offers a direct means of populating the metastable 5D$_{3/2}$ state to measure the nuclear magnetic octupole moment in the odd barium isotopes.  Light from a diode-pumped, solid state Tm,Ho:YLF laser operating at 2051 nm is used to coherently drive this transition between resolved Zeeman levels in a single trapped $^{138}$Ba$^+$ ion.  The frequency of the laser is stabilized to a high finesse Fabry Perot cavity at 1025 nm after being frequency doubled.   Rabi oscillations on this transition indicate a laser-ion coherence time of 3 ms, most likely limited by ambient magnetic field fluctuations. 
\end{abstract}


\maketitle
\section{Introduction}
The 5D$_{3/2}$ level in a barium ion has a lifetime of approximately 80 seconds \cite{dehmelt97}, corresponding to a 13 mHz linewidth for the electric quadrupole 6S$_{1/2} \leftrightarrow$ 5D$_{3/2}$ transition.  This transition is an important component of a number of proposed experiments.  First, it provides an efficient means of populating and detecting 5D$_{3/2}$ sublevels for precision RF spectroscopy of the hyperfine splittings in the  5D$_{3/2}$ manifold in $^{137}$Ba$^+$ (I=3/2), which combined with a similar measurement of the 5D$_{5/2}$ hyperfine splittings will yield the value of the nuclear magnetic octupole moment for this isotope \cite{howell08}.  This measurement could provide insight into the unexpectedly large measured value of the nuclear magnetic octupole moment of $^{133}$Cs \cite{tanner03}.  Second, the 6S$_{1/2} \leftrightarrow$ 5D$_{3/2}$ transition in $^{137}$Ba$^+$ can be used as a reference for an optical frequency standard.  In $^{137}$Ba$^+$ the quadrupole Stark shift due to electric field gradients is equal to zero for both the F=0 and F=2 hyperfine levels 5D$_{3/2}$, making $^{137}$Ba$^+$ an interesting candidate as an ion frequency standard \cite{sherman05}.  This shift had been a concern for the clock transition in $^{199}$Hg$^+$  \cite{Itano00}, but was later measured and eliminated at the 10$^{-18}$ level \cite{oskay05}.  Finally, this work is an important step towards a measurement of parity non-conservation in a single trapped barium ion \cite{fortson93}.  

Driving the 6S$_{1/2}$ to 5D$_{3/2}$ electric quadrupole transition in Ba$^+$ with high fidelity is a necessary step in each of these experiments.  We demonstrate the ability to coherently excite a $^{138}$Ba$^+$ ion from the ground state to resolved 5D$_{3/2}$ Zeeman sublevels using a frequency stabilized 2051 nm Tm,Ho:YLF laser.

\section{Apparatus}
To load ions into the trapping apparatus, barium atoms emitted from an oven are singly ionized by a two photon, isotope-selective process\cite{steele07}.  First, an external cavity diode laser (ECDL) at 791 nm drives the weak inter-combination transition, 6s$^2 \leftrightarrow$ 6s6p ($J=1$), which has isotope shifts on the order of 1 GHz between the 137 and 138 isotopes of neutral barium \cite{grundevik83}.    From this excited state, a pulse from a nitrogen laser at 337 nm excites the electron into the continuum.  Since $^{137}$Ba$^+$ represents just 11\% of a natural barium sample, this isotope-selective process allows for faster loading times for $^{137}$Ba$^+$, which will be important for future work.

 After photoionization, a single Ba$^+$ ion is then confined by a linear Paul trap in a similar design to the one found in Ref.~\cite{olmschenk07}, driven with a radio frequency (RF) of 22.9 MHz.  A pair of Helmholtz coils generate a stable, but adjustable magnetic field of up to 5 Gauss, which provides a quantization axis for the ion.  Components of the Earth's magnetic field that are perpendicular to the laboratory field are cancelled by two secondary coils.  
 
 \begin{figure}[tbp]
	\centering
		\includegraphics[]{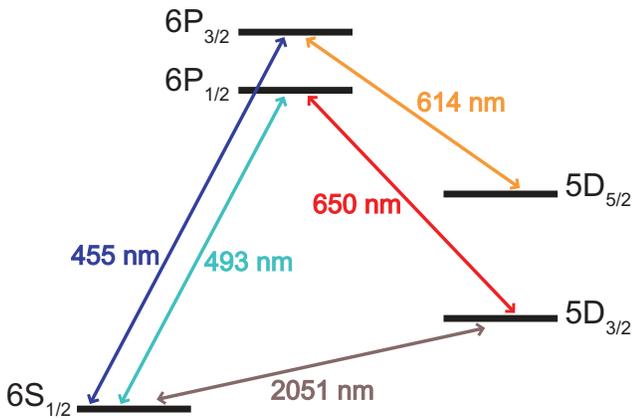}
	\caption{Energy levels and important transitions in $^{138}$Ba$^+$.  The ion is laser cooled with the 493 nm transition, and is repumped from the long lived 5D$_{3/2}$ state with 650 nm light.  The ion can be `shelved' using 455 nm light, via the 6P$_{3/2}$ state, where it will remain in the metastable 5D$_{5/2}$ state until a pulse of 614 nm light returns it to the cooling cycle. The electric quadrupole transition at 2051 nm that is of interest for proposed experiments is driven by a frequency stabilized laser, and detected by means of shelving and deshelving as described in the text.}
	\label{fig:ba_138_transitions}
\end{figure}  

The energy level diagram along with the relevant transitions for $^{138}$Ba$^+$ is shown in Fig.~\ref{fig:ba_138_transitions}. The ion is cooled to the Doppler limit T$\sim1$mK by addressing the 6S$_{1/2}\leftrightarrow$ 6P$_{1/2}$ transition at 493 nm with a frequency doubled ECDL at 986 nm.  Since this state can decay into the long lived 5D$_{3/2}$  state, the ion must be repumped using a second ECDL at 650 nm.  Both of these beams are linearly polarized for Doppler cooling. To reduce long term frequency drifts, each laser is frequency stabilized to a Zerodur-spaced optical cavity that is sealed from the atmosphere in a vacuum chamber.  Light sent to each of these cavities is frequency shifted using a double passed acousto-optic modulator (AOM) to allow for a tunable offset from the cavity modes.  The ion can be optically pumped into either of the 6S$_{1/2}$, m$_J=\pm1/2$ states by switching from a primary 493 nm beam to a second, circularly-polarized beam which is aligned parallel to the quantization axis.

Ion fluorescence from the cooling cycle is collected by a micro-objective lens, then passed through an interference filter centered at 488 nm with a transmission bandwidth of 10 nm to pass photons emitted from the 6S$_{1/2}\leftrightarrow$ 6P$_{1/2}$ transition, and block stray room light as well as scattered 650 nm photons.  These 493 nm photons are detected by a photo-multiplier tube (PMT) and imaged by an electron multiplying charged-coupled device (EMCCD) camera.  

2051 nm light is generated by a Tm,Ho:YLF diode-pumped solid-state laser manufactured by CLR Photonics (now part of Lockheed Martin) with maximum output power of 40 mW.  Changing the temperature of the resonator coarsely tunes the center wavelength of the laser.  Fine tuning of the laser frequency is achieved by adjusting the voltage applied to a piezo-electric actuator mounted on the output coupling mirror of the laser cavity, with a small-signal bandwidth of approximately 10 kHz.  This light must be frequency stabilized to a reference cavity, as detailed below, to coherently drive the narrow 6S$_{1/2} \leftrightarrow$ 5D$_{3/2}$ electric quadrupole transition. Before interacting with the ion, the 2051 nm beam passes through an AOM driven at 55 MHz (labeled `Shutter AOM' in Figure \ref{fig:schematic}), which acts as a fast shutter. 
The 2051 nm beam is linearly polarized and can be aligned parallel or perpendicular to the magnetic field so that $\Delta$m$_J=\pm1,\pm2$ transitions are allowed \cite{roos00}.   The $\Delta$m$_J=0$ transitions, which are preferred in an optical frequency standard, are not driven by light parallel or perpendicular to the magnetic field, but are maximized with light at 45 degrees to the magnetic field.  

Performing quantum jump laser spectroscopy \cite{dehmelt86} on the 6S$_{1/2}\leftrightarrow$ 5D$_{3/2}$ transition is complicated by the fact that the 2051 nm transition does not remove the ion from the cooling cycle.  It is therefore necessary to have some other means of `shelving' the ion.  A pulse from a 1W light emitting diode (LED) with a center wavelength of 455 nm and a spectral half-width of 20 nm excites an ion in the 6S$_{1/2}$ level to the 6P$_{3/2}$ level as shown in Figure~\ref{fig:ba_138_transitions}. An ion in the 6P$_{3/2}$ level will spontaneously decay back to the 6S$_{1/2}$ level with a probability of $75.6\%$, to the 5D$_{5/2}$ level with a probability of $21.5\%$, and to the 5D$_{3/2}$ level with a probability of $2.9\%$\cite{kurz08}.  A 300 ms pulse of 455 nm light effectively eliminates the possibility of the ion remaining in the 6S$_{1/2}$ level.  An ion in the 5D$_{5/2}$ level does not interact with the 493 nm and 650 nm lasers during the cooling cycle and the ion fluorescence signal at 493 nm goes to zero.  The ion can be returned to the cooling cycle by applying a 400 ms of 614 nm light that is also generated by a 1W LED with a 617 nm center wavelength, and 18 nm spectral half-width. \footnote{The LEDs in use are Phillips Luxeon III Star LEDs - Royal blue for 455 nm light, and Red-Orange for 617 nm light.}

\section{Tm,Ho:YLF Laser Frequency Stabilization}

The narrow linewidth of the 6S$_{1/2} \leftrightarrow$ 5D$_{3/2}$ transition requires that the frequency of the 2051 nm laser be well stabilized.  While reference cavities designed for use at visible and near-infrared wavelengths routinely have finesses greater than $\mathcal{F}\sim10^5$, at the time our system was built, state-of-the-art high reflective coatings at 2051 nm would be limited to $\mathcal{F}\sim20,000$.  Additionally, commercial options for electro-optic modulators (EOMs), necessary for a Pound-Drever-Hall (PDH) lock, and broadband AOMs, necessary for a wide tuning range, are not readily available at this wavelength. For these reasons we opted to stabilize the second-harmonic of our laser with a high finesse cavity at 1025 nm.

\subsection{Second Harmonic Generation of 2051 nm Light}
 Implementing a robust Pound-Drever Hall lock \cite{hall83} to a 1025 nm reference cavity requires that we generate at least 100 $\mu$W of 1025 nm light so that we can comfortably accommodate power losses from a double passed, frequency shifting AOM.   Second harmonic generation (SHG) of light is achieved using a bulk periodically poled lithium niobate (PPLN) crystal that is 4 cm long and has nine parallel tracks with different poling periods.  We used a track with a poling period of 30.25 $\mu$m and maintained the crystal temperature at $108\,^{\circ}\mathrm{C}$ for quasi-phase matching.   In single-pass configuration, however, the light generated from the PPLN nm was insufficient for a stable PDH lock. 

\begin{figure}[htbp]
\includegraphics{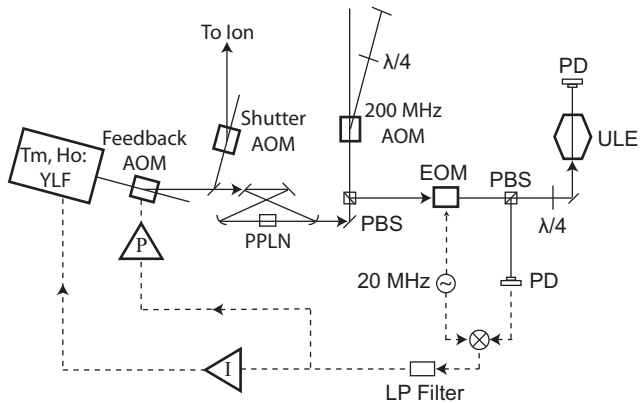}
\caption{Tm,Ho:YLF laser stabilization schematic.  Light from a 2051 nm Tm,Ho:YLF laser passes through an AOM and is frequency doubled with a PPLN crystal in a `bow-tie' cavity.  The 1025 nm light is then frequency shifted by a double-passed AOM, and stabilized to a high finesse reference cavity using the PDH method \cite{hall83}.  An EOM driven by a 20 MHz source is used to generate the sidebands necessary for frequency stabilization. After passing through a low-pass (LP) filter, high bandwidth proportional feedback (P) is sent to the feedback AOM, while low bandwidth integral feedback (I) is amplified and sent to the tuning piezo inside the laser head.}
\label{fig:schematic}
\end{figure}

1025 nm SHG efficiency is increased substantially by placing the temperature controlled PPLN crystal inside a `bow-tie' enhancement cavity (Fig.~\ref{fig:schematic}) consisting of two flat mirrors and two curved mirrors each with a radius of curvature of 250 mm.  The curved mirrors are separated by 325 mm and the PPLN crystal is placed at the midpoint where the beam comes to a 150 $\mu$m waist.  The relatively long mirror separation ensures that the beam is not clipped by the 0.5 mm square poling channels.  The crystal is anti-reflection (AR) coated for 2051 nm, as well as 1025 nm, and is been polished with a 1 degree wedge to prevent the crystal from acting as an intra-cavity etalon.  The length of the enhancement cavity is locked to the wavelength of the 2051 nm laser by sending an error signal derived using the Hansch-Couillaud method\cite{hansch80} to a piezo-electric actuator mounted on one of the flat cavity mirrors.  Our coupling efficiency into the TEM00 mode of the enhancement cavity is approximately 75\%.  We estimate the cavity build up of 2051 nm light to be approximately 20.  Sending the full 40 mW output of the Tm,Ho:YLF laser to the enhancement cavity we were able to generate 2.5 mW of 1025 nm light.  When the 55 MHz AOM (labeled `Feedback AOM') in Figure \ref{fig:schematic}, necessary for high bandwidth feedback for the PDH lock to the high finesse cavity, is included in the 2051 nm beam path, approximately 20 mW is sent to the enhancement cavity and we are able to generate approximately 500 $\mu$W of 1025 nm light.  This power is sufficient for a stable PDH lock of the laser at 2051 nm.

\subsection{PDH Stabilization with Two-Channel Feedback}
After the enhancement cavity, the 1025 nm beam makes a double pass through a frequency shifting AOM with a center frequency of 200 MHz.  This AOM allows for a continuous tuning range of approximately 80 MHz at 1025 nm (equivalent to 40 MHz of tuning at the original 2051 nm wavelength).  The beam then passes through a resonant EOM to generate Pound-Drever-Hall sidebands at 20 MHz \cite{hall83} and then through a series of mode matching lenses before coupling into the reference cavity.  

Our reference cavity consists of two mirrors held 77.5 mm apart by a spacer made from ultra-low expansion (ULE) glass according to a vibration insensitive design developed at JILA \cite{notcutt05}, resulting in a free spectral range of 1.9 GHz.  The mirrors are coated for high reflectivity at the 1025 nm wavelength.  The input mirror is flat and the output mirror has a radius of curvature of 500 mm.  Cavity ring-down measurements \cite{berden00} indicate the cavity finesse to be greater than 300,000.   The cavity is mounted vertically inside a vacuum chamber and maintained at a pressure of less than $10^{-8}$ Torr with an ion pump.  The vacuum chamber is enclosed inside a temperature stabilized aluminum box  surrounded by insulation to further reduce cavity length variations due to temperature fluctuations. 

Light reflected from the cavity is separated from the incident beam using a quarter waveplate and a polarizing beam splitter.  The intensity of the rejected beam is detected with an amplified photodetector.  The PDH error signal is extracted from the photodetector output using a double balanced mixer, a variable phase shifter, and a 1.9 MHz low-pass filter.

Frequency stabilization of the Tm,Ho:YLF laser is achieved by sending the error signal into high bandwidth and low bandwidth feedback channels.  The low bandwidth channel consists of an analog integrator circuit (labeled ``I" in Fig.~\ref{fig:schematic}) followed by a high-voltage amplification stage driving the tuning piezo inside the laser head.  The bandwidth of this channel is approximately 10 kHz.  Higher bandwidth feedback is necessary to maintain a robust lock to the $\sim$7 kHz wide TEM00 mode of the ULE cavity.  To accomplish this, the error signal in the high bandwidth channel (labeled ``P" in Fig.~\ref{fig:schematic}) is sent through an analog amplification stage that provides variable proportional gain, and then to the control input of the voltage controlled oscillator (VCO) driving the feedback AOM shown in Fig.~\ref{fig:schematic}.  Both the VCO and the AOM have 2 MHz modulation bandwidths.

\section{2051 nm Spectroscopy Procedure}
The ion is optically pumped into the 6S$_{1/2}$, m$_J=+1/2$ state with circularly polarized, 493 nm light.  The 493 nm and 650 nm lasers are then shuttered and the ion is exposed to a pulse of 2051 nm light.  The duration of the pulse is optimized to provide the maximum population transfer to the 5D$_{3/2}$ level (i.e. a Rabi $\pi$-pulse).  In order to reduce dephasing due to magnetic field fluctuations, the 2051 nm light pulse is triggered by the rising slope of the 60 Hz AC power line.  We then apply a 300 ms pulse of 455 nm light.  If the 2051 nm transition occurred, then the ion is in the 5D$_{3/2}$ level and the 455 nm light will have no effect.  If the 2051 nm transition did not occur, then the ion remains in the 6S$_{1/2}$ level and the 455 nm light will `shelve' the ion into the 5D$_{5/2}$ state, which has a natural lifetime of approximately 32 seconds\cite{kurz08}.  The 493 nm and 650 nm lasers are then turned back on and we record whether the ion fluoresces. A failed `shelving' attempt indicates a successful 2051 nm transition.

With appropriate polarization and beam alignment we can drive $\Delta$m = $\pm1,\pm2$ transitions.  To identify these transitions between Zeeman levels, we record the frequency offset from the TEM00 mode of the ULE reference cavity for each line for different magnetic field strengths created by the Helmholtz coils.  The sensitivity to a known change of the magnitude of the magnetic field can be used to identify the transitions. Driving $\Delta$m $  = \pm2$ transitions will be used to populate the sublevels needed to perform the proposed measurement of the nuclear magnetic octupole moment \cite{howell08}.

\section{Observation of 2051 nm Transitions}
\begin{figure}[htbp]
	\centering
		\includegraphics[]{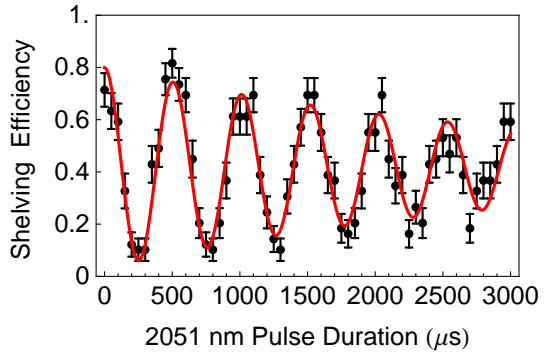}
	\caption{Coherent excitation of the 2051 nm transition.  The probability of detecting a `shelved' ion is plotted against the time for which the ion is exposed to the 2051 nm laser.  The Rabi frequency is calculated using a least-squares fit of an exponentially decaying sinusoid to be 2 kHz with a decay time of 3.2 ms.  The decoherence time is consistent with drifts in ambient magnetic fields.}
	\label{fig:rabi_data_fast}
\end{figure}

With the 2051 nm beam aligned parallel to the magnetic field, we identified and tuned the 2051 nm laser to the center of the $\Delta$m$=+1$ transition.  We then varied the duration of the 2051 nm laser pulse while recording whether the ion is `shelved'.   An optical pumping efficiency of approximately 95\% combined with imperfect `shelving' events limits the extrema of the shelving efficiency to a maximum of approximately 80\% and a minimum of 5\%.  Fitting a decaying sinusoid to the data indicates a Rabi frequency of 2.0 kHz and a decay time constant for the coherence envelope of 3.2 ms, as shown in Fig.~\ref{fig:rabi_data_fast}.  The experimental setup used in this work does not employ magnetic shielding, so the decoherence observed here is most likely caused by short-term fluctuations in the ambient magnetic field, which have been independently observed in measurements of the 6S$_{1/2}$ Zeeman splitting.

\begin{figure}[htbp]
	\centering
		\includegraphics[]{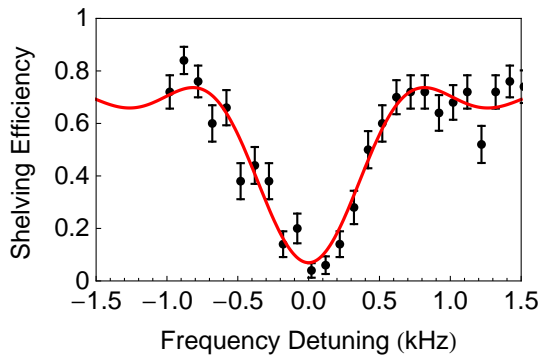}
	\caption{Spectrum of the 6S$_{1/2}$, m = +1/2 $\leftrightarrow$ 5D$_{3/2}$, m = +3/2 transition.  The probability of finding the ion in the `shelved' state is plotted against a relative frequency shift to the high finesse optical cavity's TEM00 mode.  The FWHM of this spectrum is calculated to be 700 Hz at 2051 nm, with 75\% shelving efficiency, and 95\% optical pumping efficiency.  A least-squares fit of a squared sinc function is overlaid, and was used to calculate these parameters.}
	\label{fig:2um_peak_broad}
\end{figure}

Using an attenuated 2051 nm laser beam and a pulse duration of 1 ms we observed the spectrum shown in Fig.~\ref{fig:2um_peak_broad}.  Fitting a sinc function lineshape to this peak, we find that the linewidth is Fourier transform limited by the 1 ms pulse to approximately 700 Hz.   

\begin{figure*}[h!tbp]
\centering
\includegraphics[]{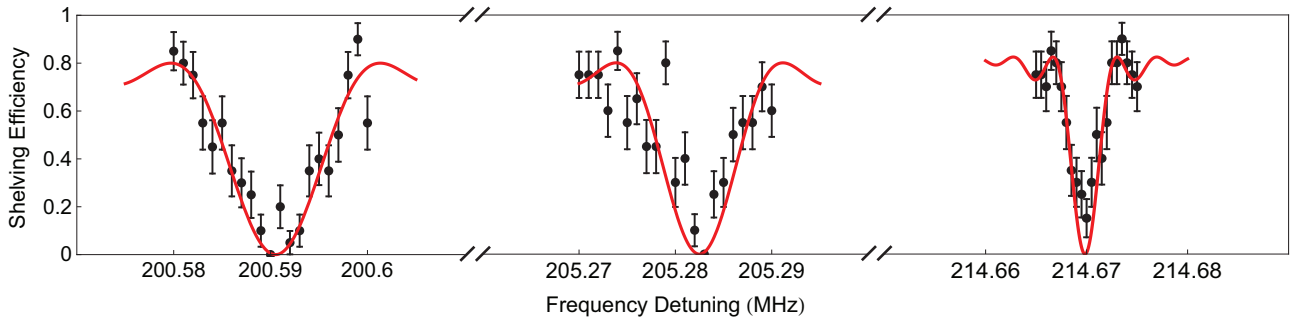}
\caption{Observation of the three accessible 6S$_{1/2}$,m=+1/2 $\leftrightarrow$ 5D$_{3/2}$ transitions using the 2051 nm laser.  The final Zeeman levels of the 5D$_{3/2}$ manifold are, from left to right,  m $=-3/2, -1/2, +3/2$.  The m=+1/2 state cannot be addressed from this initial state with our laser alignment.  The probably of finding the ion `shelved' after applying a Rabi $\pi$ pulse is plotted against a relative frequency shift to the TEM00 mode frequency of the high finesse optical cavity. A normalized sinc function (solid red line) is overlaid using the known 2051 nm laser pulse duration, as well as the overall shelving efficiency.}
\label{fig:three_plots}
\end{figure*}

With the 2051 nm beam aligned perpendicular to the magnetic field three transitions from the 6S$_{1/2}$, m = +1/2 state can be driven.  The spectra showing each of the transitions are shown in Fig.~\ref{fig:three_plots}.  The differences in frequency for these Zeeman splittings are 9.387 MHz and 4.693 MHz, so we identify the transition at 200.590 MHz as the $\Delta$m = -2 transition, the one at 205.283 MHz as the $\Delta$m=-1 transition, and the one at 214.670 MHz as the $\Delta$m = +1 transition.  Additionally, using the measured Land$\acute{\mathrm{e}}$ g-factors \cite{gfactor}, we can estimate the laboratory magnetic field to be 4.69 Gauss. The linewidths of these spectra are Fourier transform limited by Rabi $\pi$ pulses of 80, 100, and 275 $\mu$s, respectively, which correspond to the maximum Rabi frequencies observed with this configuration.

\section{Conclusions}
Using a diode-pumped solid state Tm,Ho:YLF laser at 2051 nm, we have coherently driven the 6S$_{1/2} \leftrightarrow$ 5D$_{3/2}$ transition in a single trapped $^{138}$Ba$^+$ ion.  We are able to address Zeeman levels of the 5D$_{3/2}$ manifold individually and have identified several transitions between Zeeman states.  A laser-ion coherence time of 3 ms has been observed in Rabi oscillations on one transition, which is most likely limited by ambient magnetic field noise and can be increased by adding magnetic shielding.  Having observed $\Delta$m = 2 transitions, we can proceed with the measurement of the nuclear magnetic octupole moment of $^{137}$Ba$^+$.

\begin{acknowledgments}
The authors wish to thank Amar Andalkar, Warren Nagourney, Will Trimble, and Yonatan Cohen for early work on this project, as well as helpful discussions with other members of the Blinov group, specifically Chen-Kuan Chou, Matt Dietrich, Nathan Kurz, Tom Noel, and Gang Shu.  This research was supported by National Science Foundation grant PHY-09-06494.
\end{acknowledgments}


\begin{thebibliography}{18}
\expandafter\ifx\csname natexlab\endcsname\relax\def\natexlab#1{#1}\fi
\expandafter\ifx\csname bibnamefont\endcsname\relax
  \def\bibnamefont#1{#1}\fi
\expandafter\ifx\csname bibfnamefont\endcsname\relax
  \def\bibfnamefont#1{#1}\fi
\expandafter\ifx\csname citenamefont\endcsname\relax
  \def\citenamefont#1{#1}\fi
\expandafter\ifx\csname url\endcsname\relax
  \def\url#1{\texttt{#1}}\fi
\expandafter\ifx\csname urlprefix\endcsname\relax\def\urlprefix{URL }\fi
\providecommand{\bibinfo}[2]{#2}
\providecommand{\eprint}[2][]{\url{#2}}

\bibitem[{\citenamefont{Yu et~al.}(1997)\citenamefont{Yu, Nagourney, and
  Dehmelt}}]{dehmelt97}
\bibinfo{author}{\bibfnamefont{N.}~\bibnamefont{Yu}},
  \bibinfo{author}{\bibfnamefont{W.}~\bibnamefont{Nagourney}},
  \bibnamefont{and} \bibinfo{author}{\bibfnamefont{H.}~\bibnamefont{Dehmelt}},
  \bibinfo{journal}{Phys. Rev. Lett.} \textbf{\bibinfo{volume}{78}},
  \bibinfo{pages}{4898} (\bibinfo{year}{1997}).

\bibitem[{\citenamefont{Beloy et~al.}(2008)\citenamefont{Beloy, Derevianko,
  Dzuba, Howell, Blinov, and Fortson}}]{howell08}
\bibinfo{author}{\bibfnamefont{K.}~\bibnamefont{Beloy}},
  \bibinfo{author}{\bibfnamefont{A.}~\bibnamefont{Derevianko}},
  \bibinfo{author}{\bibfnamefont{V.~A.} \bibnamefont{Dzuba}},
  \bibinfo{author}{\bibfnamefont{G.~T.} \bibnamefont{Howell}},
  \bibinfo{author}{\bibfnamefont{B.~B.} \bibnamefont{Blinov}},
  \bibnamefont{and} \bibinfo{author}{\bibfnamefont{E.~N.}
  \bibnamefont{Fortson}}, \bibinfo{journal}{Phys. Rev. A}
  \textbf{\bibinfo{volume}{77}}, \bibinfo{pages}{052503}
  (\bibinfo{year}{2008}).

\bibitem[{\citenamefont{Gerginov et~al.}(2003)\citenamefont{Gerginov,
  Derevianko, and Tanner}}]{tanner03}
\bibinfo{author}{\bibfnamefont{V.}~\bibnamefont{Gerginov}},
  \bibinfo{author}{\bibfnamefont{A.}~\bibnamefont{Derevianko}},
  \bibnamefont{and} \bibinfo{author}{\bibfnamefont{C.~E.}
  \bibnamefont{Tanner}}, \bibinfo{journal}{Phys. Rev. Lett.}
  \textbf{\bibinfo{volume}{91}}, \bibinfo{pages}{072501}
  (\bibinfo{year}{2003}).

\bibitem[{\citenamefont{Sherman et~al.}(2005)\citenamefont{Sherman, Trimble,
  Metz, Nagourney, and Fortson}}]{sherman05}
\bibinfo{author}{\bibfnamefont{J.}~\bibnamefont{Sherman}},
  \bibinfo{author}{\bibfnamefont{W.}~\bibnamefont{Trimble}},
  \bibinfo{author}{\bibfnamefont{S.}~\bibnamefont{Metz}},
  \bibinfo{author}{\bibfnamefont{W.}~\bibnamefont{Nagourney}},
  \bibnamefont{and} \bibinfo{author}{\bibfnamefont{N.}~\bibnamefont{Fortson}},
  in \emph{\bibinfo{booktitle}{LEOS Summer Topical Meetings, 2005 Digest of
  the}} (\bibinfo{year}{2005}), pp. \bibinfo{pages}{99 -- 100}, ISSN
  \bibinfo{issn}{1099-4742}.

\bibitem[{\citenamefont{{Itano}}(2000)}]{Itano00}
\bibinfo{author}{\bibfnamefont{W.}~\bibnamefont{{Itano}}},
  \bibinfo{journal}{Journal of Research of the National Institute of Standards
  and Technology} \textbf{\bibinfo{volume}{105}}, \bibinfo{pages}{829}
  (\bibinfo{year}{2000}).

\bibitem[{\citenamefont{Oskay et~al.}(2005)\citenamefont{Oskay, Itano, and
  Bergquist}}]{oskay05}
\bibinfo{author}{\bibfnamefont{W.~H.} \bibnamefont{Oskay}},
  \bibinfo{author}{\bibfnamefont{W.~M.} \bibnamefont{Itano}}, \bibnamefont{and}
  \bibinfo{author}{\bibfnamefont{J.~C.} \bibnamefont{Bergquist}},
  \bibinfo{journal}{Phys. Rev. Lett.} \textbf{\bibinfo{volume}{94}},
  \bibinfo{pages}{163001} (\bibinfo{year}{2005}).

\bibitem[{\citenamefont{Fortson}(1993)}]{fortson93}
\bibinfo{author}{\bibfnamefont{N.}~\bibnamefont{Fortson}},
  \bibinfo{journal}{Phys. Rev. Lett.} \textbf{\bibinfo{volume}{70}},
  \bibinfo{pages}{2383} (\bibinfo{year}{1993}).

\bibitem[{\citenamefont{Steele et~al.}(2007)\citenamefont{Steele, Churchill,
  Griffin, and Chapman}}]{steele07}
\bibinfo{author}{\bibfnamefont{A.~V.} \bibnamefont{Steele}},
  \bibinfo{author}{\bibfnamefont{L.~R.} \bibnamefont{Churchill}},
  \bibinfo{author}{\bibfnamefont{P.~F.} \bibnamefont{Griffin}},
  \bibnamefont{and} \bibinfo{author}{\bibfnamefont{M.~S.}
  \bibnamefont{Chapman}}, \bibinfo{journal}{Phys. Rev. A}
  \textbf{\bibinfo{volume}{75}}, \bibinfo{pages}{053404}
  (\bibinfo{year}{2007}).

\bibitem[{\citenamefont{Grundevik et~al.}(1983)\citenamefont{Grundevik,
  Gustavsson, Olsson, and Olsson}}]{grundevik83}
\bibinfo{author}{\bibfnamefont{P.}~\bibnamefont{Grundevik}},
  \bibinfo{author}{\bibfnamefont{M.}~\bibnamefont{Gustavsson}},
  \bibinfo{author}{\bibfnamefont{G.}~\bibnamefont{Olsson}}, \bibnamefont{and}
  \bibinfo{author}{\bibfnamefont{T.}~\bibnamefont{Olsson}},
  \bibinfo{journal}{Z. Phys. A} \textbf{\bibinfo{volume}{312}},
  \bibinfo{pages}{1} (\bibinfo{year}{1983}).

\bibitem[{\citenamefont{Olmschenk et~al.}(2007)\citenamefont{Olmschenk, Younge,
  Moehring, Matsukevich, Maunz, and Monroe}}]{olmschenk07}
\bibinfo{author}{\bibfnamefont{S.}~\bibnamefont{Olmschenk}},
  \bibinfo{author}{\bibfnamefont{K.~C.} \bibnamefont{Younge}},
  \bibinfo{author}{\bibfnamefont{D.~L.} \bibnamefont{Moehring}},
  \bibinfo{author}{\bibfnamefont{D.~N.} \bibnamefont{Matsukevich}},
  \bibinfo{author}{\bibfnamefont{P.}~\bibnamefont{Maunz}}, \bibnamefont{and}
  \bibinfo{author}{\bibfnamefont{C.}~\bibnamefont{Monroe}},
  \bibinfo{journal}{Phys. Rev. A} \textbf{\bibinfo{volume}{76}},
  \bibinfo{pages}{052314} (\bibinfo{year}{2007}).

\bibitem[{\citenamefont{Roos}(2000)}]{roos00}
\bibinfo{author}{\bibfnamefont{C.}~\bibnamefont{Roos}}, Ph.D. thesis,
  \bibinfo{school}{Karl-Franzens-Univ. Graz}, \bibinfo{address}{Graz}
  (\bibinfo{year}{2000}).

\bibitem[{\citenamefont{Nagourney et~al.}(1986)\citenamefont{Nagourney,
  Sandberg, and Dehmelt}}]{dehmelt86}
\bibinfo{author}{\bibfnamefont{W.}~\bibnamefont{Nagourney}},
  \bibinfo{author}{\bibfnamefont{J.}~\bibnamefont{Sandberg}}, \bibnamefont{and}
  \bibinfo{author}{\bibfnamefont{H.}~\bibnamefont{Dehmelt}},
  \bibinfo{journal}{Phys. Rev. Lett.} \textbf{\bibinfo{volume}{56}},
  \bibinfo{pages}{2797} (\bibinfo{year}{1986}).

\bibitem[{\citenamefont{Kurz et~al.}(2008)\citenamefont{Kurz, Dietrich, Shu,
  Bowler, Salacka, Mirgon, and Blinov}}]{kurz08}
\bibinfo{author}{\bibfnamefont{N.}~\bibnamefont{Kurz}},
  \bibinfo{author}{\bibfnamefont{M.~R.} \bibnamefont{Dietrich}},
  \bibinfo{author}{\bibfnamefont{G.}~\bibnamefont{Shu}},
  \bibinfo{author}{\bibfnamefont{R.}~\bibnamefont{Bowler}},
  \bibinfo{author}{\bibfnamefont{J.}~\bibnamefont{Salacka}},
  \bibinfo{author}{\bibfnamefont{V.}~\bibnamefont{Mirgon}}, \bibnamefont{and}
  \bibinfo{author}{\bibfnamefont{B.~B.} \bibnamefont{Blinov}},
  \bibinfo{journal}{Phys. Rev. A} \textbf{\bibinfo{volume}{77}},
  \bibinfo{pages}{060501} (\bibinfo{year}{2008}).

\bibitem[{\citenamefont{Drever et~al.}(1983)\citenamefont{Drever, Hall,
  Kowalski, Hough, Ford, Munley, and Ward}}]{hall83}
\bibinfo{author}{\bibfnamefont{R.~W.~P.} \bibnamefont{Drever}},
  \bibinfo{author}{\bibfnamefont{J.~L.} \bibnamefont{Hall}},
  \bibinfo{author}{\bibfnamefont{F.~V.} \bibnamefont{Kowalski}},
  \bibinfo{author}{\bibfnamefont{J.}~\bibnamefont{Hough}},
  \bibinfo{author}{\bibfnamefont{G.~M.} \bibnamefont{Ford}},
  \bibinfo{author}{\bibfnamefont{A.~J.} \bibnamefont{Munley}},
  \bibnamefont{and} \bibinfo{author}{\bibfnamefont{H.}~\bibnamefont{Ward}},
  \bibinfo{journal}{Applied Physics B: Lasers and Optics}
  \textbf{\bibinfo{volume}{31}}, \bibinfo{pages}{97} (\bibinfo{year}{1983}).

\bibitem[{\citenamefont{Hansch and Couillaud}(1980)}]{hansch80}
\bibinfo{author}{\bibfnamefont{T.}~\bibnamefont{Hansch}} \bibnamefont{and}
  \bibinfo{author}{\bibfnamefont{B.}~\bibnamefont{Couillaud}},
  \bibinfo{journal}{Optics Communications} \textbf{\bibinfo{volume}{35}},
  \bibinfo{pages}{441 } (\bibinfo{year}{1980}), ISSN \bibinfo{issn}{0030-4018}.

\bibitem[{\citenamefont{Notcutt et~al.}(2005)\citenamefont{Notcutt, Ma, Ye, and
  Hall}}]{notcutt05}
\bibinfo{author}{\bibfnamefont{M.}~\bibnamefont{Notcutt}},
  \bibinfo{author}{\bibfnamefont{L.-S.} \bibnamefont{Ma}},
  \bibinfo{author}{\bibfnamefont{J.}~\bibnamefont{Ye}}, \bibnamefont{and}
  \bibinfo{author}{\bibfnamefont{J.~L.} \bibnamefont{Hall}},
  \bibinfo{journal}{Opt. Lett.} \textbf{\bibinfo{volume}{30}},
  \bibinfo{pages}{1815} (\bibinfo{year}{2005}).

\bibitem[{\citenamefont{Berden et~al.}(2000)\citenamefont{Berden, Peeters, and
  Meijer}}]{berden00}
\bibinfo{author}{\bibfnamefont{G.}~\bibnamefont{Berden}},
  \bibinfo{author}{\bibfnamefont{R.}~\bibnamefont{Peeters}}, \bibnamefont{and}
  \bibinfo{author}{\bibfnamefont{G.}~\bibnamefont{Meijer}},
  \bibinfo{journal}{International Reviews in Physical Chemistry}
  \textbf{\bibinfo{volume}{19}}, \bibinfo{pages}{565} (\bibinfo{year}{2000}).

\bibitem[{\citenamefont{Kn\"oll et~al.}(1996)\citenamefont{Kn\"oll, Marx,
  H\"ubner, Schweikert, Stahl, Weber, and Werth}}]{gfactor}
\bibinfo{author}{\bibfnamefont{K.~H.} \bibnamefont{Kn\"oll}},
  \bibinfo{author}{\bibfnamefont{G.}~\bibnamefont{Marx}},
  \bibinfo{author}{\bibfnamefont{K.}~\bibnamefont{H\"ubner}},
  \bibinfo{author}{\bibfnamefont{F.}~\bibnamefont{Schweikert}},
  \bibinfo{author}{\bibfnamefont{S.}~\bibnamefont{Stahl}},
  \bibinfo{author}{\bibfnamefont{C.}~\bibnamefont{Weber}}, \bibnamefont{and}
  \bibinfo{author}{\bibfnamefont{G.}~\bibnamefont{Werth}},
  \bibinfo{journal}{Phys. Rev. A} \textbf{\bibinfo{volume}{54}},
  \bibinfo{pages}{1199} (\bibinfo{year}{1996}).

\end{thebibliography}
\end{document}